\documentclass[showpacs,showqkeys,amsmath,amssymb,graphicx,12pt]{revtex4}

\usepackage{graphicx}
\usepackage{epsfig}
\usepackage{multirow}
\usepackage{subfigure}
\usepackage{hyperref}
\usepackage{dcolumn}
\usepackage{bm}
\def\mop{{\langle\mathcal{O}^H_n\rangle}}
\def\OP#1#2#3#4{{\bigl.^{#1}\hspace{-1mm}{#2}_{#3}^{[#4]}}}

\def\NO{\nonumber}

\def\md{\mathrm{d}}

\def\s{\sigma}

\def\be{\begin{equation}}
\def\ee{\end{equation}}
\def\bea{\begin{eqnarray}}
\def\eea{\end{eqnarray}}
\def\bec{\begin{center}}
\def\eec{\end{center}}
\def\bei{\begin{itemize}}
\def\eei{\end{itemize}}

\def\gev{\mathrm{~GeV}}

\def\up{\Upsilon}
\def\jpsi{J/\psi}
\begin{document}

\title{FDCHQHP:A Fortran Package for Heavy Quarkonium HadroProduction}


\author{
 Lu-Ping Wan\footnote{ Corresponding author. \\ Tel: +86 01088236143. \\ E-mail addresses: wanlp@ihep.ac.cn(L.-P. Wan), jxwang@ihep.ac.cn(J.-X. Wang).
  }, Jian-Xiong Wang }%
\affiliation{
Institute of High Energy Physics, Chinese Academy Sciences, P.O. Box 918(4), Beijing, 100049, China.
}%
\date{\today}

\begin{abstract}
FDCHQHP is a Fortran package to calculate the transverse momentum($p_t$) distribution
of yield and polarization for heavy quarkonium hadroproduction at next-to-leading-order (NLO) based on non-relativistic QCD(NRQCD) framework.
It contains the complete color-singlet and color-octet intermediate states in present theoretical level, and is available to calculate different polarization parameters in different frames.
As the LHC running now and in the future, it supplies a very useful tool to obtain theoretical prediction on
the heavy quarkonium hadroproduction.

Program summary
\\~\\
\emph{Program title}:FDCHQHP v1.0 \\
\emph{Catalogue identifier}: AEBV$\_$v1$\_$0 \\
\emph{Program summary URL}:\href{http://v-www.ihep.ac.cn/~wjx/genpack/genpack.html}{http://v-www.ihep.ac.cn/~wjx/genpack/genpack.html} \\
\emph{Program obtainable from}: CPC Program Library, Queens University, Belfast, N. Ireland \\
\emph{Licensing provisions}: Standard CPC licence, http://cpc.cs.qub.ac.uk/licence/licence.html \\
\emph{No. of lines in distributed program, including test data, etc.}: 3 199 202 \\
\emph{No. of bytes in distributed program, including test data, etc.}: 592M \\
\emph{Distribution format}: taz \\
\emph{Programming language}: Fortran 77 \\
\emph{Computer}: Any computer with Linux operating system and Intel Fortran Compiler \\
\emph{Operating system}: Linux  \\
\emph{Classification}: 11.1 \\
\emph{Nature of problem}: This package is for the calculation of the heavy quarkonium hadroproduction at NRQCD NLO. \\
\emph{Solution method}: The Fortran codes of this package are generated by FDC system [1] automatically.  \\
\emph{Additional comments}: It is better to run the package in supercomputers.  \\
\emph{Running time}: For a independent sub-process, it may cost several seconds to several hours depending on the number of sample points if one CPU core is used.
For a complete prompt production of heavy quarkonium hadroproduction ($\psi(Ns)$ or $\Upsilon(Ns)$) at one $p_t$ point, it may cost one hour to thousands of hours depending on the number of sample points  if one CPU core is used. In our test with less sample points, it take us 16 minutes to compile and 25 minutes to run the whole program  by 4 processes with the CPU $Intel^{\circledR}$ Core. I7-3770k @ 3.5GHz $\times$ 8 .\\
\end{abstract}

\maketitle

\section{Introduction}

From the discovery of heavy quarkonium in the 1970s, both the theoretical and experimental study on
production and decay of $J/\psi$ and $\Upsilon$, plays an important role in the research on the nonperturbative and
perturbative aspects of QCD. The factorization framework of non-relativistic QCD
(NRQCD) had been proposed to study the production and decay of heavy quarkonium~\cite{Bodwin:1994jh}.
By extracting the NRQCD long-distance matrix elements (LDMEs) from the matching between theoretical prediction and
experimental data, the NRQCD calculation had given a good description on the transverse momentum distribution ($p_t$)
of heavy quarkonium production at hadron colliders. But the polarization of heavy quarkonium hadroproduction had been
an open question for many years.

In the past, many works has been done by applying the Fortran codes generated by our Feynman Diagram Calculation(FDC)
package~\cite{Wang:2004du} which is written in REDUCE and RLisp and can generate Fortran codes automatically.
In the recent two works~\cite{Gong:2012ug, Gong:2013qka}, we have presented the complete next-to-leading-order(NLO)
predictions to the $p_t$ distribution of yield and polarization for $J/\psi$ and $\up$ prompt hadroproduction. It took us a lot
of efforts to manipulate running of more than ten thousands the numerical calculation jobs and control the final numerical
precision when more than hundred numerical results from sub-processes must be sum up for each polarization result at each $p_t$ point.
Finally, based on the collection of all the Fortran codes generated by using FDC, the efforts and experience on job submission
and numerical precision control are implemented in
an independent package FDCHQHP, it can be easily run by inexperienced user to obtain any desirable results on heavy quarkonium
hadroproduction at any given experimental condition.
It is well known that only prompt experimental measurements on heavy quarkonium hadroproduction are available up to now.
Therefore, to fit the experimental measurement, the theoretical prediction must be on the prompt production.
FDCHQHP supplys the complete calculation on the $p_t$ distribution of yield and polarization for prompt hadroproduction,
which includes direct production part and feed-down contribution form excited states production.

From another aspect,  as the LHC running at the current energy, the $p_t$ distribution of yield are measured up
to large $p_t$ region (e.g. the measurement of $\up$ by ATLAS are up to 70GeV ~\cite{Aad:2012dlq}), and there are
more polarization parameters measurement e.g. $\lambda_\theta$, $\lambda_{\theta\phi}$, $\lambda_{\phi}$ in different
reference frame for heavy quarkonium polarization, and also the
frame-independent polarization parameter $\widetilde{\lambda}$ ~\cite{Chatrchyan:2012woa}.
What's more, in the next years, the LHC running energy will upgrade to 14 TeV, and there will be
more high-precision measurements to be published in the future. Although we have already finished a few theoretical prediction
in comparison with the experimental measurements in past two work. There are still a lot of experimental measurements
in future need the theoretical predictions which needed large scale calculation in supercomputer environment. There are already
many request for theoretical prediction in different experimental condition, we are ready to make the package public so
that it can be used to obtained theoretical prediction by other people.

Based on NRQCD formalism, the cross section of $h$ hadroproduction is
\bea
\s[pp\rightarrow hx]=\sum\int dx_1 dx_2 G^i_pG^j_p \hat{\s}[ij\rightarrow (c\bar{c})_nx]\mop,
\eea \
where $p$ is either a proton or antiproton,
the indices $i, j$ run over all the partonic species and $n$ represents the $c\bar{c}$ intermediate states
($\OP{3}{S}{1}{1}$, $\OP{3}{S}{1}{8}$, $\OP{1}{S}{0}{8}$, $\OP{3}{P}{J}{8}$) for $\jpsi$ and $\psi(2s)$,
or ($\OP{3}{P}{J}{1}$, $\OP{3}{S}{1}{8}$) for $\chi_{cJ}$.
The short-distance contribution  $\hat{\s}$ can be perturbatively calculated and the long-distance matrix elements (LDMEs) $\mop$ represent the nonperturbative QCD effects.
FDCHQHP package contains all the sub-processes via color-singlet(CS) and color-octet(CO) intermediate states up to the NLO
level for heavy quarkonium prompt production.  We list all the sub-processes in Tab~\ref{tab:tab-feyn}.

\begin{table}[ht]
\scalebox{0.8}{
\begin{tabular}{c|c|c|c|c}
\hline
STATES&  LO sub-process  &  number of         &  NLO sub-process  &   number of  \\[0.5ex]
      &                  &  Feynman diagrams  &                   &   Feynman diagrams \\[0.5ex] \hline
$^3S_1^{1}$ &  $g+g\rightarrow$$(Q\bar{Q})_n+g$  &  6  & $g+g\rightarrow$$(Q\bar{Q})_n+g$(one-loop)  & 128\\[0.5ex] \cline{4-5}
&  &  &  $g+g\rightarrow$$(Q\bar{Q})_n+g+g$ &  60  \\[0.5ex] \cline{4-5}
&  &  &  $g+g\rightarrow$$(Q\bar{Q})_n+Q+\bar{Q}$  &  42  \\[0.5ex] \cline{4-5}
&  &  & $g+g\rightarrow$$(Q\bar{Q})_n+q+\bar{q}$ &  6  \\[0.5ex] \cline{4-5}
&  &   &  $g+q(\bar{q})\rightarrow$$(Q\bar{Q})_n+g+q(\bar{q})$  &  6  \\[0.5ex] \cline{4-5}
\hline
$^1S_0^{8}$(also $^3P_J^{8}$) &  $g+g\rightarrow$$(Q\bar{Q})_n+g$  &  (12,16,12)  & $g+g\rightarrow$$(Q\bar{Q})_n+g$(one-loop)  & (369,644,390)\\[0.5ex]
\cline{2-5}
or &  $g+q(\bar{q})\rightarrow$$(Q\bar{Q})_n+q(\bar{q})$ & (2,5,2) & $g+q(\bar{q})\rightarrow$$(Q\bar{Q})_n+q(\bar{q})$(one-loop) & (61,156,65)\\[0.5ex] \cline{2-5}
$^3S_1^{8}$&  $q+\bar{q}\rightarrow$$(Q\bar{Q})_n+g$  & (2,5,2)  &  $q+\bar{q}\rightarrow$$(Q\bar{Q})_n+g$(one-loop)  & (61,156,65)\\[0.5ex] \cline{2-5}
or &  &  &  $g+g\rightarrow$$(Q\bar{Q})_n+g+g$ &  (98,123,98)  \\[0.5ex] \cline{4-5}
$^3P_J^{1}$ &  &  &  $g+g\rightarrow$$(Q\bar{Q})_n+q+\bar{q}$ &  (20,36,20)  \\[0.5ex] \cline{4-5}
  &  &  &  $g+q(\bar{q})\rightarrow$$(Q\bar{Q})_n+g+q(\bar{q})$  &  (20,36,20)  \\[0.5ex] \cline{4-5}
&  &  &  $q+\bar{q}\rightarrow$$(Q\bar{Q})_n+g+g$  &  (20,36,20)  \\[0.5ex] \cline{4-5}
  &  &  &  $q+\bar{q}\rightarrow$$(Q\bar{Q})_n+q+\bar{q}$ &  (4,14,4)  \\[0.5ex] \cline{4-5}
 &  &  &  $q+\bar{q}\rightarrow$$(Q\bar{Q})_n+q^{\prime}+\bar{q^{\prime}}$ &  (2,7,2) \\[0.5ex] \cline{4-5}
 &  &  &  $q+q\rightarrow$$(Q\bar{Q})_n+q+q$ &  (4,14,4)  \\[0.5ex] \cline{4-5}
 &  &  &  $q+q^{\prime}\rightarrow$$(Q\bar{Q})_n+q+q^{\prime}$ &  (2,7,2)  \\[0.5ex] \cline{4-5}
\hline
\end{tabular}
}
\caption{
The sub-processes for heavy quarkonium $c\bar{c}$ and $b\bar{b}$ prompt production at LO and NLO, and the number of the
corresponding Feynman diagrams.  The numbers in the round brackets of the third and fifth columns, denote the number of
Feynman diagrams for the intermediate states $^1S_0^{8}$(also $^3P_J^{8}$),$^3S_1^{8}$(for both the direct part and the $\chi_{c(b)}$ state),
$^3P_J^{1}$( for $\chi_{c(b)}$ state) from left to right, respectively. In total, there are almost 5,000 Feynman diagrams.
}
\label{tab:tab-feyn}
\end{table}

For FDCHQHP, besides the yield of heavy quarkonium, the most important feature is to calculate
the polarization parameters $\lambda_\theta$, $\lambda_{\theta\phi}$, $\lambda_{\phi}$, which is defined as
in Ref.~\cite{Beneke:1998re}:
\be
\lambda_\theta=\frac{\md\sigma_{11}-\md\sigma_{00}}{\md\sigma_{11}+\md\sigma_{00}}, \lambda_{\theta\phi}=\frac{\sqrt{2}\mathrm{Re}\md\sigma_{10}}{\md\sigma_{11}+\md\sigma_{00}},
\lambda_{\phi}=\frac{2\md\sigma_{1,-1}}{\md\sigma_{11}+\md\sigma_{00}}, \NO
\ee
where $d\sigma_{S_zS_z^\prime}$ is the spin density matrix of heavy quarkonium hadroproduction.
Therefore, the $p_t$ distribution of yield is obtained as $\md\sigma=2\md\sigma_{11}+\md\sigma_{00}$.

This package includes 6 channels and 76 sub-processes as summarized in above table, almost 2 millions lines Fortran codes in total.
It can be run in complete paralleled mode with more than thousands of cpu cores with very high efficiency,
For test, it can be run on the personal computer. It is supposed to be run on supercomputer for real application
since it is really CPU-time consuming calculation.  It provide a program WBIN (written in bash-shell script) to run
at supercomputer, to do job submission and numerical precision control automatically.
Until now, we have used FDCHQHP on two supercomputers in China: one is shenteng7000 provided by CNIC, CAS;
the other is TH-1A provided by NSCC-TJ.

\section{Installation}

The FDCHQHP can be run under Linux system, with Fortran compiler and MPI library, and part of the Fortran source
needs the quadruple precision support. Intel Fortran compiler ifort and the MPI library are needed.
The necessary FDC library and BASES library~\cite{Kawabata:1995th,Yuasa:1999an} are included in the package for convenience.
The packages are distributed in a compressed file named FDCHQHP.taz and can be downloaded from FDC homepage:

\href{http://v-www.ihep.ac.cn/~wjx/genpack/genpack.html}{http://v-www.ihep.ac.cn/~wjx/genpack/genpack.html}

The installation steps are written in shell scripts to simplify the installation.
Users can install the package as following:

\bei
\item Place FDCHQHP.taz into a directory and decompress it with   \\
      \textbf{tar -xzvf FDCHQHP.taz} ,\\
      then the home directory \emph{FDCHQHP} is generated.
\item Set the environmental variables \lq fdc\rq  with the absolute path where the directory \emph{FDCHQHP/fdc2.0} exists,
      which \lq fdc\rq  is used in the next compiling process.
      If TC shell is used and \emph{FDCHQHP} is placed in the home directory, users can add the below line in  $\sim$/\emph{.cshrc}: \\
      \textbf{setenv fdc $\$$HOME/FDCHQHP/fdc2.0 }
\item  Compile the libraries in the \emph{fdc2.0} folder\\
       Before compiling, it is necessary to check the compiler used in the file \emph{fdc-build}. It could be \lq mpif77\rq
       or \lq mpiifort\rq  or \lq ifort\rq  for Intel Fortran Compiler.
       And then execute \\
      \textbf{./fdc-build} \\
\item Compile the Fortran codes in the \emph{source} folder \\
      Users also should check the compiler used in the file \emph{source-make}, and then execute  \\
     \textbf{ ./source-make}  \\
      For the first time, users can use the above command to do a entirely compiling automatically,
      it took us about 16 minutes to finish it with 4 CPU cores in our personal computer($Intel^{\circledR}$ Core. I7-3770k @ 3.5GHz $\times$ 8.
      It is also available for users to go into a certain sub-process and compile the codes again if the codes are partly changed.
      Please note that, if the file \emph{makefile} was changed, you should execute \textbf{make clean} first, and then execute
      \textbf{make}, otherwise, just \textbf{make} is OK.
\eei

\section{The directory structures and programs}
\subsection{ Structure of directories  }
All the Fortran codes are collected in the folder \emph{FDCHQHP}.
The folder's structure is like :\\
\bei
\item  \emph{source/$<$channels$>$/$<$sub-processes$>$/fort/}
\item  \emph{fdc2.0/f77/src/src64d\_ifort(src64d\_ifort\_p100, src64q\_ifort)/}
\item  \emph{fdc2.0/basesv5.1/basesMPI/}
\item  \emph{fdc2.0/basesv5.1/lib/}
\eei

In \emph{source}, the six $<$channels$>$ are named as \emph{1s0\_8, 3pj\_8, 3s1\_1, 3s1\_8, chic-3pj\_8, chic-3s1\_8} , to represent
the physical channels $^1S_0^8, ^3P_J^8, ^3S_1^1, ^3S_1^8, \chi_{c(b)J}(^3P_J^1), \chi_{c(b)J}(^3S_1^8)$, respectively.
The $<$sub-processes$>$ under each $<$channel$>$ are the ones listed in Tab.~\ref{tab:tab-feyn}, they are named with
\lq g\rq or  \lq q\rq, which denotes the partons in initial and final states except the heavy quarkonium.
For example, \emph{gggg} denotes $g+g\rightarrow J/\psi(\up)+g+g$, \emph{qggq} denotes $g+q(\bar{q})\rightarrow J/\psi(\up)+g+q(\bar{q})$,
and so on. Without specification of quadruple precision, all the calculation are double precision calculation.
Especially, in the channel named \emph{3pj\_8}, the sub-process \emph{ggg\_loop32}, denotes the one-loop part for
$g+g\rightarrow J/\psi(\up)+g$(one-loop) with quadruple precision calculation Fortran source involved.
A corresponding sub-process, \emph{ggg\_other} denotes the remaining part aside from the loop part(i.e. \emph{ggg\_loop32} )
with double precision calculation.

We know that the physical processes are divided into several channels, and a channel are divided into a few sub-processes.
Therefore, the final result in physics should be the summation of all of them.
For any sub-processes, the Fortran source codes and the corresponding compiled executable program (i.e. \emph{int})
are placed in the \emph{fort} directory.

\subsection{Programs }
In the \emph{fort} directory, there are several kinds of files, and the most important one for users is \emph{input.dat} which
present all the input to run the Fortran program.  We describe them in the following.

\subsubsection{\emph{Introduction to the file \emph{input.dat}}}
\begin{table}[ht]
\scalebox{0.8}{
\begin{tabular}{c c|c}
\cline{1-3}
 \textbf{input.dat} & \ \  & \ \ \ \textbf{introduction} \ \ \  \\[0.5ex] \cline{1-3}
      1       &     number\_of\_beam1 &   number of the first beam energys \\[0.5ex] \cline{1-3}
     3500    &     e\_beam1           &   first beam's energy, $\gev$ \\[0.5ex] \cline{1-3}
    1      &     number\_of\_beam2    &   number of the first beam energys \\[0.5ex] \cline{1-3}
     3500    &     e\_beam2           &   second beam's energy, $\gev$  \\[0.5ex] \cline{1-3}
     -4.0d0   &     y1min             &   rapidity cut for the heavy quarkonium\\[0.5ex] \cline{1-2}
     -2.5d0   &     y1max             &                   \\[0.5ex] \cline{1-2}
     2.5d0    &     y2min             &     \         \\[0.5ex] \cline{1-2}
     4.0d0    &     y2max             &            \\[0.5ex] \cline{1-3}
     1.5      &     mc                &    heavy quark mass, $\gev$\\[0.5ex] \cline{1-3}
     0.81    &     Rs\_0$\hat{\ }$2   &    square of the original wave function for $^3S_1^1$, $\gev^3$  \ \  \\[0.5ex] \cline{1-3}
    0.012   &     O\_8\_1\_\_0        &    LDME for $^1S_0^8$ state, $\gev^3$ \\[0.5ex] \cline{1-3}
     0.0039   &     O\_8\_3\_\_1      &   LDME for $^3S_1^8$ sate, $\gev^3$ \\[0.5ex] \cline{1-3}
     0.027   &     O\_8\_3\_P\_J      &   LDME for $^3P_J^8$ sate, $\gev^5$ \\[0.5ex] \cline{1-3}
     0.075     &    Rs$\prime{}$\_chic\_0$\hat{\ }$2  &   square of the derivative for wave
                   function at original point $\chi_{cJ}^{^3P_J^1}$ part, $\gev^5$  \\[0.5ex]\cline{1-3}
    0.012     &    Rs\_chic\_0$\hat{\ }$2*Br0  &  the branch ratio for \\ [0.5ex] \cline{1-2}
    0.344     &    Rs\_chic\_1$\hat{\ }$2*Br0  &  $\chi_{cJ}$ feed-down to $J/\psi$    \\[0.5ex] \cline{1-2}
    0.195     &    Rs\_chic\_2$\hat{\ }$2*Br0  &  when J=0,1,2, respectively   \\ [0.5ex]  \cline{1-3}

     0.001    &     scut              &   soft cut in the phase space two-cutoffs method \\[0.5ex] \cline{1-3}
     0.00002  &     ccut              &   co-line divergence cut in the phase space two-cutoffs method \\[0.5ex] \cline{1-3}
    3        &     nf                &   number of light quarks \\[0.5ex] \cline{1-3}
      1       &     c\_miu\_r        &   ratio of $\mu_r \over \mu_0$, $\mu_r$ is the renormalization scale,
                    $\mu_0=\sqrt{4m_q^2+p_t^2}$.  \\[0.5ex] \cline{1-3}
     1        &     c\_miu\_f        &   ratio of $\mu_f \over \mu_0$, $\mu_f$ is the factorization scale,
                    $\mu_0=\sqrt{4m_q^2+p_t^2}$. \\[0.5ex] \cline{1-3}
     4       &    polar\_axis\_0\_helicity\_1\_CS\_  &  Polarization frame: 0 helicity frame, 1 Collins-Soper frame, \\
             &     2\_Target\_3\_GJ\_4\_Recoil  &~ ~ ~ ~ ~ ~ ~ 2 target frame, 3 GJ frame, 4 Recoil frame \\ [0.5ex] \cline{1-3}
    1     &  me2\_type\_1\_11\_\_2\_1-1\_\_3\_10\_\_4\_00    &  Polarization density matrix: 1 transverse rate ${d\sigma_{1,1}\over dp_t}
             (={d\sigma_{-1,-1}\over dp_t}$) ,4 longitude rate ${d\sigma_{0,0}\over dp_t}$,\\
        &     & 2 and 3 are the non-diagonal ones ${d\sigma_{1,-1}\over dp_t}, {d\sigma_{1,0}\over dp_t}$ , respectively.
             $nb/\gev$ \\[0.5ex] \cline{1-3}
     1       &     cncall &    a float number between  0.01 to $10^3$, the number of MC integration sample points\\
             &            &    is set as {\bf cncall} multiple the initial number of sample point in the file int.f \\[0.5ex] \cline{1-3}
    1        &     ncoll\_1\_lhc\_2\_tevatron     &   1 LHC, 2 Tevatron, with the corresponding PDF configuation\\[0.5ex] \cline{1-3}
    21       &     number\_of\_pt\_points     &    the number of $p_t$ points \\[0.5ex] \cline{1-3}
     3.0      &                          &   the $p_t$ data, $\gev$ \\ [0.5ex]
   $\vdots$     &                    &   their total number should coincide with \\[0.5ex]
    50.0     &            &  number\_of\_pt\_points above  \\[0.5ex] \cline{1-3}

\end{tabular}}
\caption{ Each line in file \emph{input.dat} are shown in the tablet. They are all the input parameters for the calculation on
the $p_t$ distribution of transverse polarized  direct $J/\psi$ hadroproduction in the experimental condition at 7 TeV LHC
with rapidity cut $2.5<|y|<4.0$.}
\label{tab-input-direct}
\end{table}

For the direct production part, including $^1S_0^8, ^3P_J^8, ^3S_1^1, ^3S_1^8$ ,
as well as for the feed-down contributions from $\chi_{c}$ in the directories \emph{chic-3pj\_8} and \emph{chic-3s1\_8},
the file \emph{input.dat} are completely same, just as Tab.\ref{tab-input-direct}.

As an example, when we calculated the $p_t$ distribution of polarization parameter($\alpha$) for $J/\psi$ prompt
production under a certain experimental measurement (here denotes the rapidity cut) at the LHC~\cite{Gong:2012ug},
we prepared two \emph{input.dat} files which uses the format showing in Tab.\ref{tab-input-direct} with me2\_type
set as 4 for longitude polarization and 1 for transverse polarization;
Other parameters can be easily changed according to the detailed description in Tab.\ref{tab-input-direct}.

\subsubsection{\emph{Introduction to other files}}

Users can skip over this block directly, if they do not concern the details of other files.
\bei
\item The files prefixed with \emph{amp}, are the amplitude calculation part�� \\
the \emph{ampN.f} is the amplitude for the Nth tree-diagram, \emph{amplN.f} is that for the Nth loop-diagram, \emph{amplNn2.f,
amplNn3.f and amplNn4.f} are the ${1 \over \varepsilon_{IR}}$, ${1 \over \varepsilon^2_{IR}}$ and ${1 \over \varepsilon_{UV}}$ divergence
terms of the Nth loop-diagram, respectively.

In \emph{amps2.f} file, it calls all the diagrams contribution and squares the matrix elements, there are different blocks for
the tree-diagram part, the loop-diagram part, the remaining term from divergence canceling part.
It also contains the different choices for polarization direction and frame.
\item
The files prefixed with \emph{incl} contain the common blocks and the definition of data, used in the Fortran programs.
\item
\emph{makefile}, list the rules, libraries with location to link and compiler for building the executable file \emph{int}.
\item
\emph{parameter.f} and \emph{parameter1.f} are two files defining the physical parameters in calculation.
\item
\emph{func.f}, generate a certain phase space point with a set of random number and calculate the cross section at the point.
\item
\emph{int.f}, the main program, read all input and calls the Monte Carlo program for phase space integration.
\item
\emph{genppp.f}, and \emph{coupling.f} calculate all the Lorentz invariable and constants in the programs.
\item
\emph{cteq6m.tbl}, or \emph{cteq6l.tbl}, are the Parton Distribution Function(PDF)~\cite{Pumplin:2002vw} used.
\item
\emph{bases.data} and the files prefixed with \emph{plot}, are generated automatically in calculation, which are usually not used.
\item
\emph{input.dat}, lists the input parameters and provides an interface for users. \\
We make two tables in this paper to explain it as Tab. \ref{tab-input-direct}.
\item
The files \emph{fresult.dat} and \emph{convergence.dat}, are the results we expect after programs running.
The former is used in the data processing, and the latter is the precision analysis for the integration.
They will be replaced by new ones if the executable \textbf{int} runs again.
We will give some detailed introduction in the section \textbf{OUTPUT FILES}.
\item
Other files not listed.
\eei

\section{usage}

\subsection{Calculation}
As we know, a complete prompt result should be the sum up of all the channels with fixed LDMEs.
Users can also do the calculation for one sub-process independently.
The following is the calculation introduction.

\subsubsection{ calculation for a sub-process}
 Once the programs compiled successfully, the executable file \emph{int} will be found in the \emph{fort} folder.
Two calculation methods are available:
   \bei
\item \textbf{./int}
\item \textbf{mpirun -n X ./int} ( X is the number of CPU cores).
   \eei
It may take a few seconds to a few hours depending on the process if just one CPU is used. Usually, the loop diagrams will
take more time.  No matter which method is used, the input.dat and cteq6m.tbl(or other PDF used) files should be placed
in the calculating folders correctly.

\subsubsection{ calculation for a channel or channels}
Go into the \emph{source} directory, execute \\
\textbf{./run-all} \\
 to do the calculation for all channel.
 It may take users several hours to a few days depending on the calculation precision and computer.

By modifying the variable \$op in line 4 with any channel name in the file \emph{run-all}, such as $op='3s1\_1/*/fort'$,
 users can obtain the calculation result for the specified channel.

In the script \emph{run-all}, we have processed the data collecting step.
The results for different channels are listed in the corresponding folders, named as \emph{data\_*}.

\section{Large scale calculation in supercomputers }
To simplify the tedious manual work, we have developed a bash-shell script package WBIN to do the numerical calculation automatically
on supercomputers.
This package provides a very simply interface and some commands for users.
It has been realized to automatically parallize the calculation, generate and submit jobs,  analyze the accuracy of the results,
fix the suitable number of CPU to each job, separately calculate the positive and negative part of a sub-process when needed,
realize the test calculation(for 3 $p_t$ points) and the real calculation(for all $p_t$ points), check the state of each job,
and finally collect all the data.
It is customized for our package, but also could be adapted in any similar application which use BASES MC integral program.

\subsection{Installation}

The package can be downloaded together with the Fortran package, named as \emph{WBIN.taz}.

There are two steps to install it:
\bei
\item  execute \textbf{tar -xzvf WBIN.taz} ;
\item  execute \textbf{cd WBIN/}
\item  add the absolute directory of WBIN to the default search path, e.g. add the line \lq set path=(\$path \$HOME/WBIN)\rq ~in \emph{$\sim$/.cshrc} .
\eei
\subsection{Usage}
In the calculation of transverse momentum distribution of yield and polarization parameters of heavy quarkonium hadroproduction,
all the results used the $p_t$ as the their x-coordinate and there are usually many $p_t$ points.
The result at these $p_t$ points can be calculated independently. Instead of the sequential calculation mode for
all the $p_t$ points at our personal computer environment, the calculation could be parallelled for all
he $p_t$ points at the same time in supercomputer environment. Therefore,
the parallelization scale will become very large dramatically. Finally, there are many jobs have to be submitted to finished
in the parallel computing environment. And each job is for calculation on one polarization component, one sub-process at one $p_t$
point with one experimental condition.

We need to do a three $p_t$ points test running with low accuracy requirement for the result
of each channel which is obtained by summing up all the sub-directories in this channel.  In the {\bf test calculation}, the number
of sample point in the Monte Carlo integration is adjusted to obtained the required accuracy, as well as the number of CPU
cores is adjusted to keep the calculation finish at two hours.
With all the information obtained in the successful {\bf test calculation}, we can arrange how to run program in each sub-process
so that the accuracy requirement for the final result of each channel can be achieved. For each jobs, the number of
CUP cores and number of sample point in the Monte Carlo integration are fixed by applying the information obtained in
{\bf test calculation} so that every jobs will be finished within two hours by themselves. Then the results for all
$p_t$ points can be obtained in the {\bf real calculation} stage.

\subsubsection{preparation}
The most important preparation is the method of submitting jobs, which depends on the job management system in supercomputers.
In WBIN package, two files refer to this method: the file \emph{Submit} and \emph{sub\_job}.
It needs users themselves to understand the method of submitting jobs in their supercomputer environment to modify it properly.
Here we provide two examples which was applied by us at the Shenteng7000 and TH-1A in China.
\bei
\item For Shenteng7000, we give the example as files \emph{Submit\_shenteng} and \emph{sub\_job\_shenteng}. \\
It adapts the LSF(Load Sharing Facility) HPC7.0 job management system at the Shenteng7000.
There is no upper limit for the total number of CPU cores applied by users, but every job submitted is added in task queues
together with other users' jobs.
\item For TH-1A, we give the example as files \emph{Submit\_th-1a} and \emph{sub\_job\_th-1a}. \\
In TH-1A, it adapts the method of batch jobs.
We can apply a certain number of CPU nodes within a upper limit, in these CPU nodes, we can arrange our small scale jobs.
Once the CPU nodes are allocated to a user, the user's jobs will monopolize the CPU nodes until all the jobs are over.
It is called a non-preemptive scheduling system.
\eei

\subsubsection{\bf  test calculation}
\begin{figure}
\center{
\includegraphics*[scale=0.8]{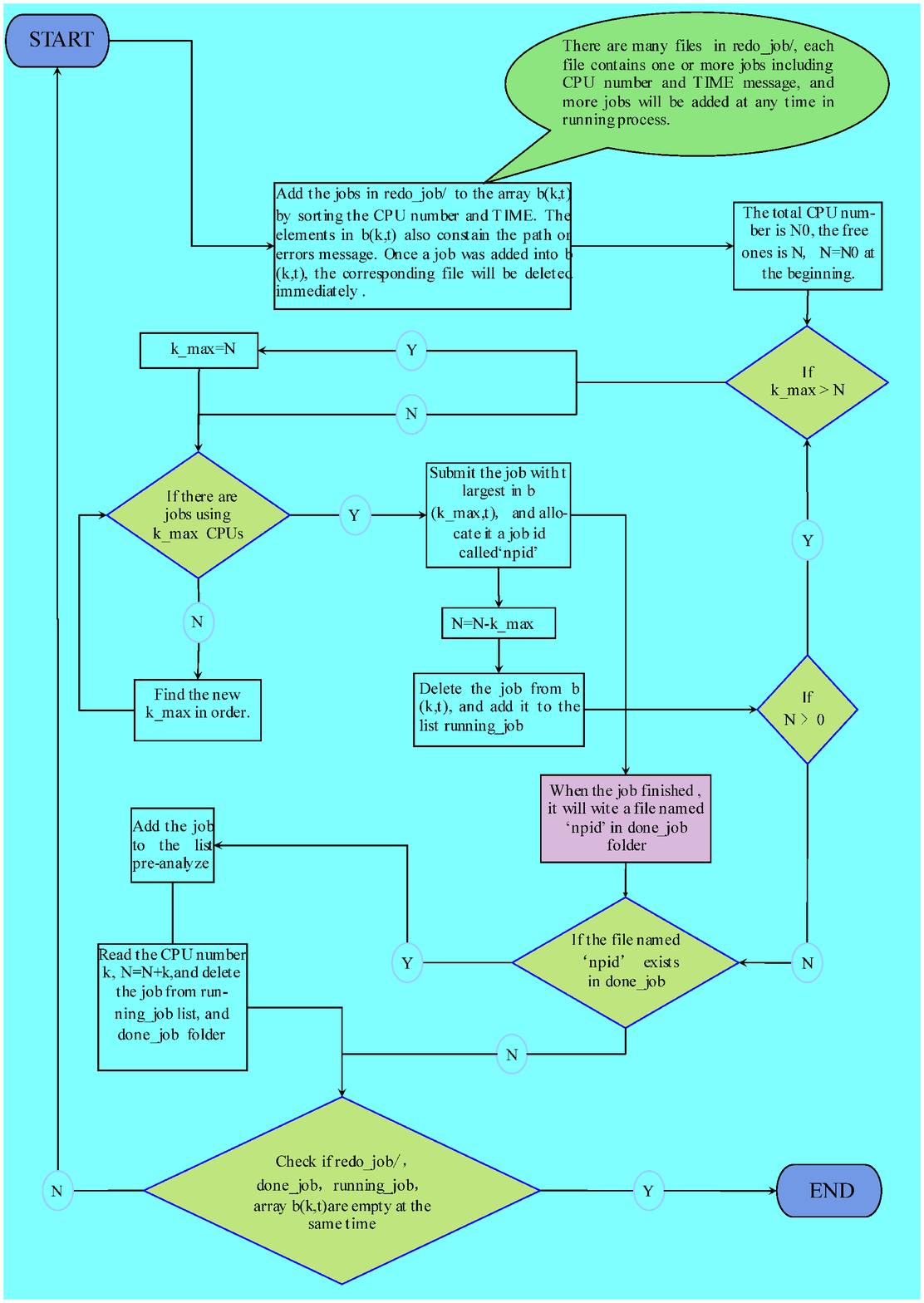}
\caption {\label{fig:flow-sub} The flow chart for submitting jobs on supercomputers in WBIN package.}}
\end{figure}

For the usage of WBIN package, these steps will be done by manual operation or by shell script as the following shows.
\bei
\item {\bf A:}  Make sure all of the sub-processes in folder \emph{FDCHQHP/source} have been compiled successfully.
\item {\bf B:}  Go into the directory \emph{source}, execute \textbf{echo \$PWD $>$ dir\_s}.
\item {\bf C:}  Make a directory any where which is specially used to do the calculation, e.g. \textbf{mkdir $\sim$/work}.
\item {\bf D:}  Copy the files \emph{all-pwd, dir\_s, input.dat, cteq6m.tbl} in \emph{source} folder into the folder \emph{work}.
\item {\bf E:}  Go into the folder \emph{work}, and use the template file \emph{input.dat} to generate more input files. \\
 In details, copy \emph{input.dat} to other files named as \emph{inputX.dat} with X can be any letters except \lq.\rq, and then modify \emph{inputX.dat} with the options you want according to Tab.~\ref{tab-input-direct}.
  The number of \emph{inputX.dat} is not limited, such as \emph{inputlp.dat, inputtp.dat, inputdirectlp.dat ...}.
\item {\bf F:}  If users want to calculate part of the total 76 sub-processes, the user can just keep the corresponding lines and delete others
in file \emph{all-pwd}.
\item {\bf G:}  Execute \textbf{init\_test\_job}. The flow for this step: \\
0)  generate the folders \emph{redo\_job, done\_job} in \emph{work}; \\
1)  generate the directories according to \emph{inputX.dat}, all the X value will be the directory name in the folder \emph{work};  \\
2)  generate the multiple sub-directories in directory \emph{X} according to the file \emph{all-pwd},  such as \emph{X/1s0\_8/gggg/},
and copy the \emph{inputX.dat} file into \emph{X/1s0\_8/gggg/input.dat}; \\
3) generate the sub-directories in \emph{X/1s0\_8/gggg/} according to $p_t$ points in the current \emph{input.dat} and name them as \emph{tedir1, tedir2, tedir3...,} for test calculation,
and then generate \emph{input.dat} in \emph{tedir1, tedir2, tedir3...,} with just one $p_t$ point as the $p_t$ list at the end of \emph{input.dat}; \\
4) link the file \emph{cteq6m.tbl} in the \emph{work} folder to \emph{tedir1, tedir2, tedir3...,}; \\
5) add a job in the file \emph{work/redo\_job/job\_list}; \\
6) repeat the steps 3), 4), 5) until the jobs for each $p_t$ point in \emph{X/1s0\_8/gggg/input.dat} has been added in \emph{work/redo\_job/job\_list}; \\
7) repeat 2)-6) for each similar sub-directory; \\
8) repeat 1)-7) for each \emph{inputX.dat}.
\item {\bf H:} Submit jobs with executing \textbf{Submit}. \\
It will submit all the jobs listed in the file \emph{work/redo\_job/job\_list}. Each line in the file denotes a job and looks like: \\
\textbf{1:1:exp1/dire/lp/1s0\_8/gggg/tedir1:0:1:1:1:wwww:6:5:0:0:3:3:int} \\
and they correspond to the variables in used by the program as\\
\textbf{Nodes:TIM:dir:Tno:err0:no\_pt:ol:fname:max\_ncpu\_job:max\_Tno \\
:kk1:kk2:max\_kk1:max\_kk2:diff\_int}. \\
The command \textbf{Submit} will read the file \emph{work/redo\_job/job\_list} line by line and stores the data in memory
in the form of arrays.
The flow chart for submitting jobs is shown in Fig.~\ref{fig:flow-sub}.
\item {\bf I:} When all the jobs submitted in step {\bf H:} have been finished, users can execute \textbf{re\_cal\_test} to judge
whether the jobs are finished successfully or aborted by check the file \emph{work/redo\_job/job\_list}.
If the file \emph{work/redo\_job/job\_list} is empty, it means that all the jobs are finished successfully, otherwise it means that
the jobs in the file \emph{work/redo\_job/job\_list} are aborted.
For the later case, users can execute \textbf{Submit} for another time just as step {\bf H:} does.
\item  {\bf J:} Repeat the steps {\bf H:} and {\bf I:} until \emph{work/redo\_job/job\_list} is empty.
\eei

\subsubsection{\bf real calculation}
The above steps are the complete flow for the test calculation(only calculate three $p_t$ points), and the flow
for real calculation(calculate all the $p_t$ points ) is almost the same with a little differences: \\
a) Ignore the steps {\bf A:-F:}. \\
b) In step {\bf G:}, execute \textbf{init\_real\_job} instead of \textbf{init\_test\_job}.
Here the folder for each $p_t$ point is named as \emph{redir3.0, redir5.35... }, where the number in the names denote the $p_t$ points. \\
b1) In addition, execute \emph{cpu\_time\_cal} to get the message of the total estimated run-time for the whole real calculation .\\
c) Execute \textbf{Submit} as the same as step {\bf H:}. \\
d) In step {\bf I:}, execute \textbf{re\_real\_job} instead of \textbf{re\_test\_job}. \\
e) The step {\bf J:} is also working for real calculation.  \\

\subsubsection{data collection}
The test and real calculations make up the whole process for the calculation in supercomputers.
Finally, execute \textbf{final\_data\_collect} to collect the data for all the $p_t$ points in all the sub-processes, and then
give the final results for the 6 channels.

\subsection{some detailed processing }
\subsubsection{control the calculation precision}
\begin{figure}
\center{
\includegraphics*[scale=0.8]{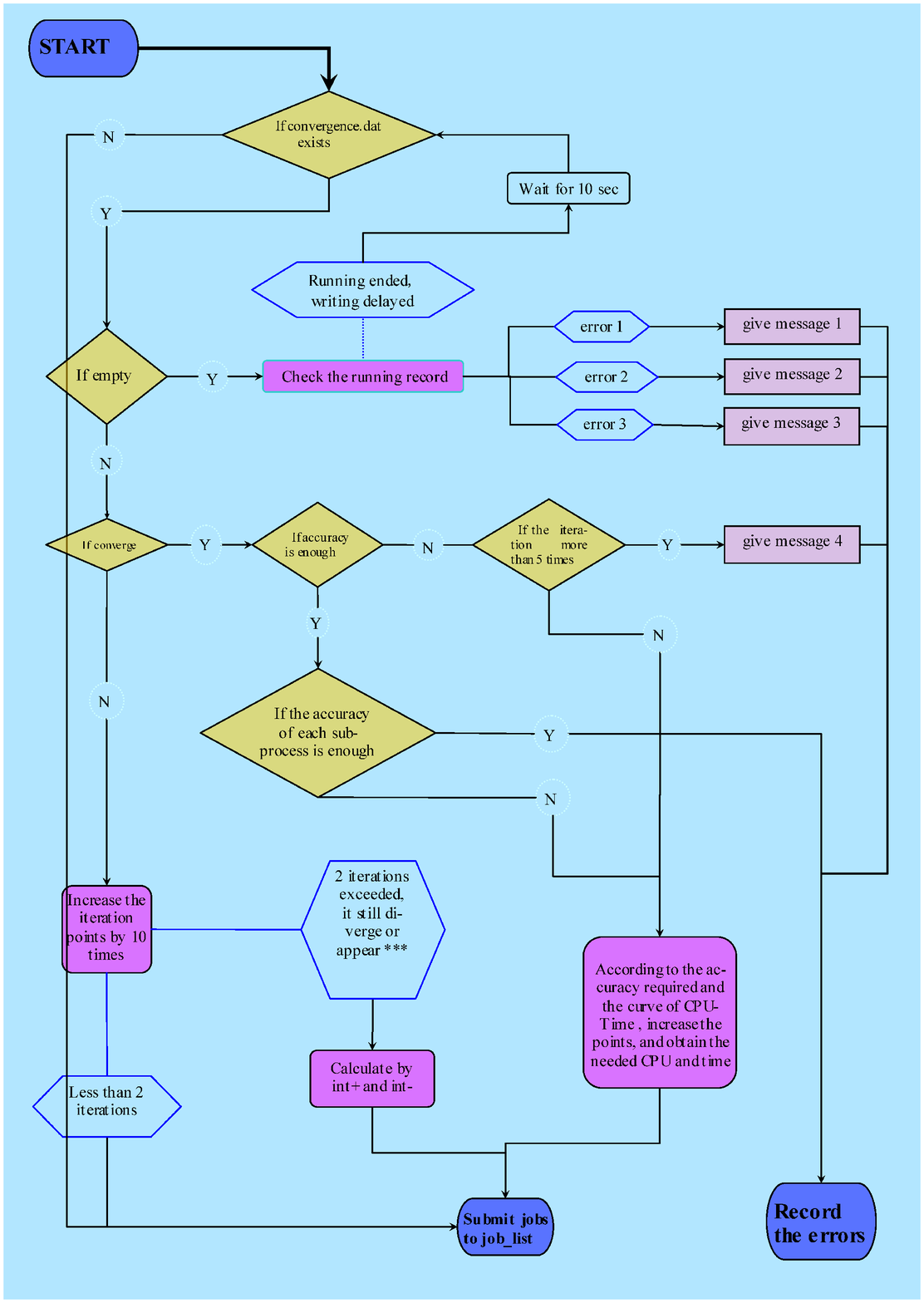}
\caption {\label{fig:flow-anly} The flow chart for analyzing the precision in WBIN package.}}
\end{figure}

Using the Monte Carlo integral program,  the error estimate and the square root of integral points are in inverse ratio.
Therefore, if we know the error for a certain number of sample points, it's easy to obtain the error for another number
of sample points.
\be
C=\sum_{i=1}^Nc_i,  E=\sqrt{\sum_{i=1}^Ne_i^2},  R=\frac{E}{|C|},   r_i=\frac{e_i}{|c_i|} ,
\ee
where N is the total number of sub-processes in a channel, $c_i$ is the central value for each sub-process, $e_i$ is
the absolute error for each sub-process, and $r_i$ is the relative error for each sub-process.
Therefore R is the relative error for a channel.
Consuming the maximal relative error for a channel we desired is $R_0$(the default value is 1\%), when $R > R_0$, we can
easily obtain that in each sub-process, the relative error should be
\be
r_i^\prime=\frac{R_0 \times C}{c_i \times \sqrt{N}}.
\ee
Using the new error $r_i^\prime$, the error $r_i$ and the number of sample points $Ncall$ in the last calculation,
we can get the new number of sample points $Ncall^\prime$:
 \be
 Ncall^\prime={(\frac{r_i}{r_i^\prime})}^2 \times Ncall .
 \ee
$Ncall^\prime$ is the new number of sample points in the next calculation process.
It may do several iterations until $R < R_0$ is achieved.
This flow chart of these steps is shown in Fig.~\ref{fig:flow-anly}.

The calculation includes two steps.
 The first step is {\bf test calculation}, in which we set a less precise error value $r_i=1\%$ for each sub-process, then
do the calculation only at three $p_t$ points which are the minimal, intermediate, and the maximal number of $p_t$ points
listed in \emph{input.dat}.
In this step, the relative error $R$ maybe not meet the precision requirement for the result of a channel
in the case of large number cancellation among different sub-processes in this channel.
Then another iteration will work until $R$ is small enough just as mentioned above.
The second step is {\bf real calculation}, in which all the $p_t$ points listed in \emph{input.dat} will be calculated.
Using the method of linear continuation for the number of sample points at the 3 $p_t$ points calculated in test step,
we can obtain the numbers of sample points for all other $p_t$ points listed in \emph{input.dat}.
In this case, the calculation precision in the whole $p_t$ region could be guaranteed $R_0$.

\subsubsection{other details}
\bei
\item In some sub-processes, for the $p_t$ points in the region where the central value changes from positive to negative or in inverse,
the precision will get worse in such region.
In this case, we adapt to do the calculation for positive and negative part, respectively, and then sum up these two parts as
the central value for the sub-process.
\item  In each sub-process, the number of sample points decides the run-time, so it is easy to estimate the number of CPU cores
and the time needed in large number of sample points through the information obtained in small sample number case.
\item  When using the same number of CPU cores, run-time for different sub-processes are quite different.
So we need to allocate the suitable number of CPU cores for each job to keep it can be finished in two hours.
\item In WBIN package, we decide the priority to submit job
by the number of CPU cores needed and then estimated run-time for the jobs.
\eei

\subsubsection{a brief guide for the shell scripts}
You can skip this subsection if you just want to run the package. It is presented here just for user who want
to involved in modification of program in the package.
\bei
\item Initialize the directories and the job\_list in test calculation---~\emph{init\_test\_job, get\_test\_pt}
\item Job Controlling---~\emph{job\_control}
\item Submit the jobs---~\emph{sub\_job}
\item Analyze the running state of job---~\emph{anly\_job}
\item Analyze the precision---~\emph{anly\_precision}
\item Separately calculate the positive and negative part---~\emph{cal\_int+-, check\_int+-\_err}
\item Check data---~\emph{check\_pro\_(test, real), check\_chan\_(test, real)}
\item Recalculate while interrupted---~\emph{re\_cal\_test, re\_cal\_real}
\item Shift from test to real calculation---~\emph{col\_test\_data, test\_to\_real, init\_real\_job}
\item Other small scripts
\eei

\section{data processing}
In FDCHQHP package , the result for a channel ``n'' are summing up of all the sub-processes in this channel as:
\be
\hat{\s}[ij\rightarrow (Q\bar{Q})_nk]=\sum^a\sigma^a_n+ 2(1-ln\frac{M_H}{\mu_\Lambda}) \times \sum^{b}\sigma^{b}_n ,
\label{eqn:processsum}
\ee
where i, j, k denotes the partons, $\sigma^a_n$ is the result in the sub-process (corresponding to a) directories
without the suffix '-m' in folder names, and $\sigma^{b}_n$ is the result in the sub-process directories with '-m'
suffix in folder names.  The second part in the Eq.(\ref{eqn:processsum}) only appears in the channels \emph{3pj\_8}
and \emph{chic-3pj\_1}. It depends on the choice of $\mu_\Lambda$ which usually take the value of $m_Q$.
Users can take different $\mu_\Lambda$ values to study the dependence on NRQCD scheme, as we did in Ref.~\cite{Gong:2013qka}.

Considering the kinematics effect for the feed-down contribution from high excited states, a shift
$p_t^H \approx p_t^{H^\prime}\times(M_H/M_{H^\prime})$ could be performed after obtaining the result of a whole channel.

\section{output files and the precision analysis}
\begin{figure}
\center{
\includegraphics[scale=1]{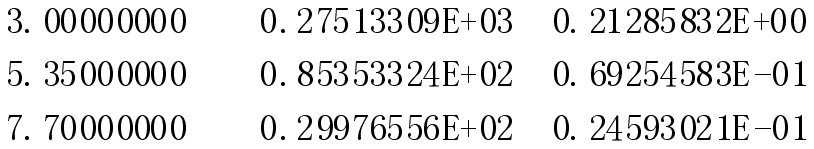}
\caption {
\label{fig:fre}
 The format of \emph{fresult.dat}. In each line, the first column is the $p_t$ point, in GeV units; the second is the central value,
in nb/GeV units;
and the third is the absolute error.
}
}
\end{figure}

\begin{figure}
\center{
\includegraphics*[scale=0.8]{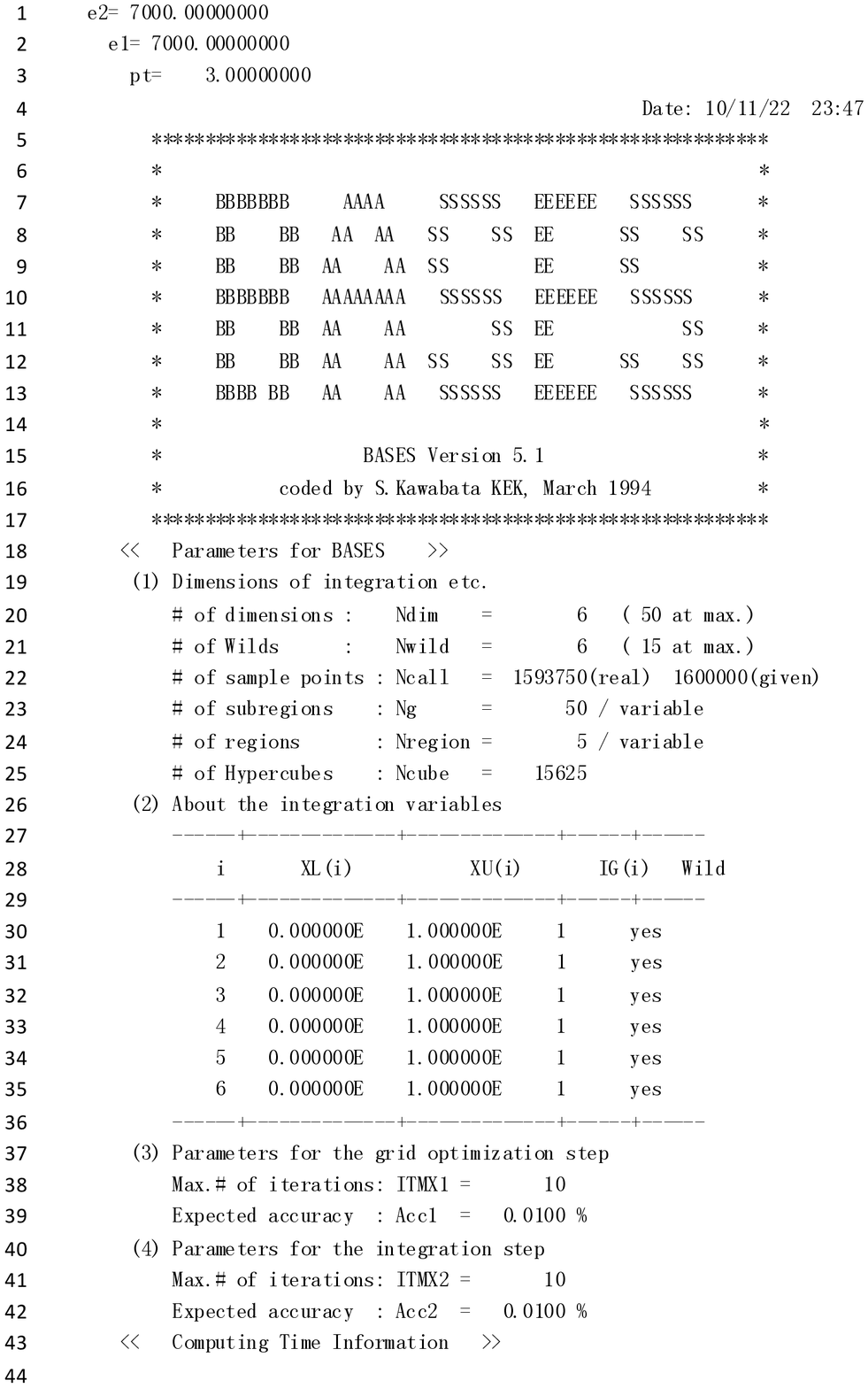}
\caption {\label{fig:conv1} The first half of convergence.dat.}}
\end{figure}

\begin{figure}
\center{
\includegraphics*[scale=0.8]{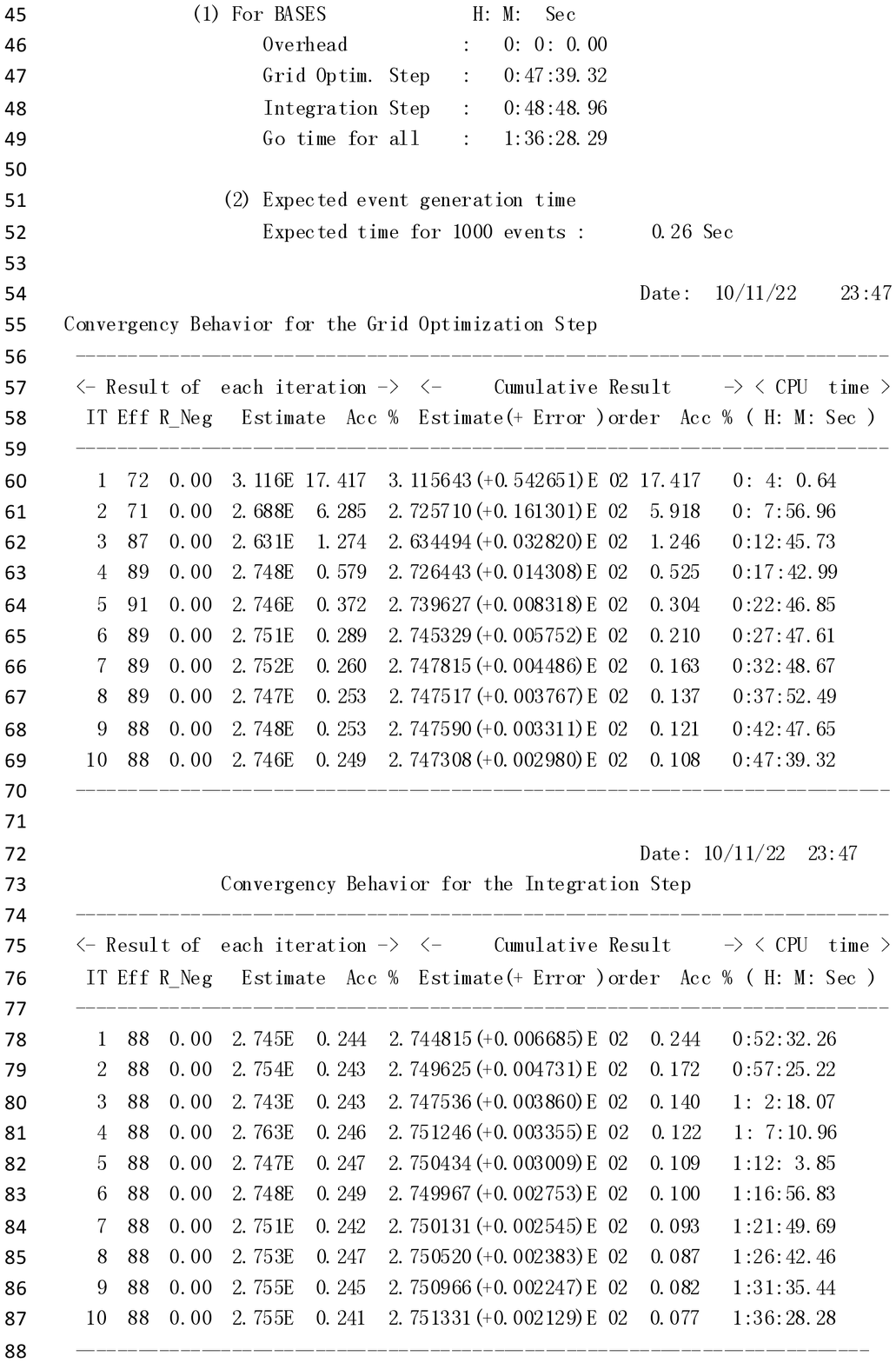}
\caption {\label{fig:conv2} The latter half of convergence.dat.}}
\end{figure}

As mentioned above, the files \emph{fresult.dat} and \emph{convergence.dat} are the output files needed.
The format of\emph{fresult.dat} is shown in Fig.~\ref{fig:fre}, and the format of \emph{convergence.dat} are shown in Fig.\ref{fig:conv1} and Fig.\ref{fig:conv2}.



The following are some notes for \emph{convergence.dat}:
\bei
\item
In Fig.\ref{fig:conv1}, the first two lines are the beam energies for colliders; the 3rd line is the $p_t$ point, the 22nd shows the number of sample points (given and real used).
\item
The 38th and 41st lines show the maximal times for the grid optimization step and integration step,
and the 39th and 42nd lines give the highest accuracy we expected in the two steps, usually they are set to be 0.01\%.
\item
In Fig.\ref{fig:conv2}, the lines 45th to 49th lines give a statistical run-time for the calculation on one $p_t$ point.
The lines from 60th to 69th are the result of each iteration in the grid optimization step. From left to right, they are
the iteration number, the efficiency, the portion of the negative value, integration value, the relative error,
the cumulative result for the data on the front(the absolute error), the total relative error, the time when each iteration finished.
\item
Lines 78th to 87th, are the reliable result in the integration step.
In this step, after the grid optimization step, the results for each iteration are close to each other,
and the final cumulative central value is the average of these data,
the corresponding relative error(e.g. the 8 column of line 78, which is 0.077\%) is the 1/$\sqrt{10}$ times
as the first line(e.g. the 8th column of line 87, 0.244\% ).
\eei

There are two standards to ensure that the convergence behavior of integration is good enough: \\
~~~~1), the relative error in the last 10 lines of convergence.dat is average and small enough, no more than 1\%
(if the accuracy is as high as almost 0.01\%, the less average result also could be accepted); \\
~~~~2), on the base of 1) achieved, the final cumulative accuracy in the last line, which is also the accuracy presented in \emph{fresult.dat} in the form of absolute error, should meet the required precision when all the sub-processes are summed, since we know that large number cancellation usually appears in the summation of the sub-processes. \\
In fact, we also require that the relative error for each sub-process itself is no more than 1\%( when it is larger than 1\% ,
the relative error in each iteration must be more than 3\%, that means the the integration is not reliable).
Besides, for a certain sub-process, when some parameters are changed, such as the rapidity range, the $p_t$ points,
the polarization direction, and so on, the convergence behavior of integration could change at the same time.
In this case, the number of sample points should be reset for the sub-processes to ensure a smooth result in the whole $p_t$ region,
especially for the loop-diagram sub-processes such as \lq $g+g\rightarrow$$J/\psi+g$ \rq, and the sub-process
\lq $g+g\rightarrow$$J/\psi+g+g$ \rq,
since the value of these two processes are usually large and opposite, which could lead to large number canceling.

Fig.\ref{fig:conv1} and Fig.\ref{fig:conv2} is a complete result for  one $p_t$ point.
If more $p_t$ points were listed in input.dat file, the results will be written into one convergence.dat in the order of $p_t$ points,
as well as the file fresult.dat.

\section{Summary}
We have presented the Fortran package FDCHQHP, which is used to do the calculation on the $p_t$ distribution of
yield and polarization for heavy quarkonium  prompt production in hadron colliders.
The package contains all the sub-processes of CS and CO states up to NLO level, and it focus on the calculation
of the polarization parameters.

This package has been fully developed in our past works with rather reliable performance and accuracy.
We also provide a small package WBIN, which can help users to run FDCHQHP in supercomputers easily.
Some other improvement are in progress so that we can make it more automatically.

\section{Acknowledgments}

We thank Dr. Bin Gong for the discussions, and also thank for the help from the Deepcomp7000 project of the
Supercomputing Center, CNIC, CAS and  TH-1A project of NSCC-TJ.
This work is supported, in part, by the National Natural Science Foundation of China (No. 10935012, and No. 11005137),
DFG and NSFC (CRC110), and by CAS under Project No. INFO-115-B01.

\bibliography{paper}
\end{document}